\documentclass[]{emulateapj}

\begin{document}


\title{The Star-Formation History of BCGs to \lowercase{z} = 1.8 from the SpARCS/SWIRE Survey: Evidence for significant in-situ star formation at high-redshift}


\author{Tracy M.A. Webb\altaffilmark{1} }
%\affil{Astronomy Department, University of California,
  %  Berkeley, CA 94720}

%\author{C. D. Biemesderfer\altaffilmark{4,5}}
%\affil{National Optical Astronomy Observatories, Tucson, AZ 85719}
%\email{aastex-help@aas.org}

%\and
\author{Adam Muzzin\altaffilmark{2}}
\author{Allison Noble\altaffilmark{6}}

\author{Nina Bonaventura\altaffilmark{1}}
\author{James Geach\altaffilmark{4}}
\author{Yashar Hezevah\altaffilmark{3}}
\author{Chris Lidman\altaffilmark{5}}
\author{Gillian Wilson\altaffilmark{7}}
\author{H.K.C. Yee\altaffilmark{6}}
\author{Jason Surace\altaffilmark{8}}
\author{David Shupe\altaffilmark{9}}

%\affil{Space Telescope Science Institute, Baltimore, MD 21218}


\altaffiltext{1}{McGill University, 3600 rue University, Montreal, QC, Canada, H3A 2T8}
\altaffiltext{2}{Leiden Observatory, University of Leiden, PO Box 9514, 2300 RA Leiden, The Netherlands}

\altaffiltext{3}{Kavli Institue for Particle Physics and Cosmology, Stanford University, 452 Lomita Mall, Stanford, CA 94305-4085}
\altaffiltext{4}{Centre for Astrophysics Research, University of Hertfordshire, Hatfield, Hertfordshire, AL109AB, UK}

\altaffiltext{5}{Australian Astronomical Observatory, PO Box 915, North Ryde, NSW 1670, Australia}

\altaffiltext{6}{Department of Astronomy and Astrophysics, University of Toronto, 50 St.~George St., Toronto, ON, Canada, M5S 3H4 }
%\altaffiltext{4}{Cape Town}
%\altaffiltext{5}{University of Chicago

\altaffiltext{7}{Department of Physics and Astronomy, University of California,  Riverside, CA,  92521, USA}
\altaffiltext{8}{Spitzer Science Center, California Institute of Technology, M/S 314-6, Pasadena, CA, 91125, USA}
\altaffiltext{9}{NASA Herschel Science Center, IPAC, 770 South Wilson Avenue, Pasadena, CA, 91125, USA}
%\altaffiltext{8}{New York State?}
%\altaffiltext{2}{Society of Fellows, Harvard University.}
%\altaffiltext{3}{present address: Center for Astrophysics,
%    60 Garden Street, Cambridge, MA 02138}
%\altaffiltext{4}{Visiting Programmer, Space Telescope Science Institute}
%\altaffiltext{5}{Patron, Alonso's Bar and Grill}


\begin{abstract}
We present the  results of a MIPS-24$\mu$m study of the Brightest Cluster Galaxies (BCGs)  of 535 high-redshift galaxy clusters.  The clusters are drawn from the Spitzer Adaptation of the Red-Sequence Cluster Survey (SpARCS), which  effectively provides a sample selected on total stellar mass,  over  0.2 $<z<$ 1.8 within the {\it Spitzer} Wide-Area Infrared Extragalactic (SWIRE) Survey  fields.  20\%, or 106 clusters have spectroscopically confirmed redshifts, and the rest have redshifts estimated from the color of their red sequence.  A comparison with the public SWIRE images  detects  125 individual BCGs at 24$\mu$m $\gtrsim$ 100$\mu$Jy, or 23\%.   The luminosity-limited detection rate of BCGs in similar richness clusters ($N_\mathrm{gal}>$ 12) increases rapidly with redshift. Above $z\sim$ 1, an average of $\sim$ 20\% of the sample have 24$\mu$m-inferred infrared luminosities of $L_{IR} >$ 10$^{12}$ $L_\odot$, while the fraction below $z\sim$ 1 exhibiting such   luminosities is $<$ 1 \%.  The {\it Spitzer}-IRAC  colors  indicate the bulk  of the 24$\mu$m-detected population  is predominantly powered by star formation, with only 7/125 galaxies lying within the color region inhabited by Active Galactic Nuclei (AGN).   Simple arguments limit the  star-formation activity to several hundred million years and this may therefore be indicative of the timescale for  AGN feedback to halt the star formation.  
Below redshift $z\sim$ 1 there is not enough star formation to significantly contribute to the overall stellar mass of the BCG population, and therefore BCG growth is likely dominated by  dry-mergers. Above $z\sim$ 1, however, the inferred star formation would double the stellar mass of the BCGs and is  comparable to the mass assembly predicted by simulations through dry mergers.  We cannot yet constrain the process driving the star formation for the overall sample, though a single object studied in detail is consistent with a gas-rich merger.

%Though the observational limits make it difficult to characterize the magnitude of the IR  evolution within BCG population, it  appears steeper than is seen for infrared field galaxies and  therefore supports the idea that different physical processes drive the evolution of central galaxies. 
%Previous studies have invoked cooling flows, however  the SpARCS sample does not have uniform or complete X-ray coverage. We therefore characterize the BCGs through their mid-infrared colors and  find the sample to be a heterogenous composition of (in order of fraction of the entire sample): infrared-faint early-type galaxies,  infrared-bright AGN, star-bursting systems,  and, importantly, strong gravitational lenses.   In the last case, the 6$''$ beam of Spitzer-MIPS means the infrared flux cannot be uniquely assigned to the BCG or the background system.  MORE 

\end{abstract}





\section{Introduction}

At the centre of  most local galaxy clusters lies a single massive galaxy.  These so-called Brightest Cluster Galaxies (BCGs) are the most massive galaxies in the universe today.  They do not appear, however, to simply be the extremes of the local galaxy mass function, but are separate beasts, with luminosities, metallicities and surface brightness profiles that are unique and consistent across the population  \citep{oemler76,tremaine77,dressler78,schombert86}.  It seems likely that the formation of BCGs is tied to the overall growth of their parent clusters, through the physics of gas cooling in the most massive dark matter halos and their hierarchical accretion of the general field galaxy population. 

Nevertheless, we know rather little about the formation histories of BCGs.  Recent measurements by \citet{lidman12} show that BCGs have increased their stellar mass by 2$\times$ since $z \sim$ 1 \citep[c.f][]{collins09,stott11}.  These authors find the growth is driven by  dry accretion of  satellite galaxies, in line with  Semi-Analytic Models \citep{bower06,delucia07,guo10}.  The accreted  systems are gas-poor and no new star formation is induced during the merger process.  In this picture most of the stellar mass is formed at very high redshifts ($z\sim$ 5) within multiple galaxies through low levels of star formation, and the BCG slowly acquires its identity through the conglomeration of previously assembled pieces.  On the other hand, several moderate redshift BCGs ($z<0.6$) exhibit signs of star formation (10-1000 $M_\odot$yr$^{-1}$) and/or contain  large amounts of molecular gas  \citep{johnstone87,mcnamara92,allen92,cardiel98,crawford99,edge01,wilman06,edwards07,odea10,donahue11,rawle12,mcdonald12}, indicating the scenario is not so simple.  In many cases the star formation rates are correlated with the gas cooling time, and indeed star formation has been primarily seen in cool-core clusters.  It is not yet clear how important  this star formation is  to the overall growth of BCGs as a population, nor how it might relate to similar processes at very high redshift  that are posited by the models.

 Star formation  fed by rapid gas deposition and continual dry  galaxy accretion are two very different scenarios of BCG evolution, and each have  implications for the physics of halo collapse and growth.  They are not  mutually exclusive  since both  are transient events regulated by different processes and may occur at different times in the history of a galaxy.  BCG galaxy accretion is  an ongoing, though sporadic, phenomenon, fed by the continual in-fall of field galaxies into the cluster potential, which are  stripped of much of  their interstellar medium (ISM) during their  descent into the centre of the cluster.     Cooling flows, and the resulting central starburst, likely have defined duty-cycles that are governed by still unclear heating and cooling mechanisms \citep{mcnamara07,rafferty08,voit08}

Studies of galaxy clusters and their central galaxies suffer from observational bias and, until recently, small sample sizes. Much of the work on central star formation has been carried out on X-ray selected clusters which  may preferentially select cooling flow clusters. Quite recently, large samples of Sunyaev-Zeldovich clusters have become available through facilities such as the South Pole Telescope \citep[e.g.][]{mcdonald13a}, that also select clusters through observations (though indirect) of their intercluster medium. 
Here we investigate the observed IR properties of BCGs using a large sample  of   clusters from the Spitzer Adaptation of the Red-Sequence Cluster Survey (SpARCS),  an optical/NIR galaxy-selected cluster survey, designed to discover  clusters to $z\sim$  2 \citep{muzzin09,wilson09}. 
While an important expansion into new parameter space, this approach also introduces new limitations and challenges. With a large sample of galaxy clusters  considerably less is known about each individual system and the conclusions are generally statistical.  It is necessary to automate some aspects of the analysis which increases the risk of sample contamination. 
Thus, we present the results of this work with the reminder that ongoing studies, such as high-resolution imaging and additional spectroscopy, are required to better calibrate some of the assumptions made here, and relate the findings to those of X-ray and SZ-selected samples.

The paper is outlined as follows.
In \S 2 we introduce the SpARCS cluster sample.  In \S 3 we outline our BCG identification algorithm and the IR analysis using the SWIRE data. In  \S 4 we discuss the origin of the IR emission and the change in the IR properties of BCGs with redshift. 
\S 5 finishes with  a discussion of the implications of \S4 and we  present our conclusions in \S6.  Standard cosmology (H$_\circ$ = 70 km/s/Mpc; $\Omega_\mathrm{matter}$ = 0.3; $\Omega_\Lambda$ = 0.7) is adopted throughout. 

\section{Data and Observations}

\subsection{The Spitzer Adaptation of the Red-Sequence Cluster Survey}

The Spitzer Adaptation of the Red-sequence Cluster Survey (SpARCS) is a deep $z'$-band imaging survey covering the {\it Spitzer} SWIRE Legacy fields, designed to assemble a  large homogeneously-selected sample of massive clusters  to $z\sim$2 \citep{muzzin09,wilson09,demarco10,muzzin12}. It employs an infrared adaptation of the cluster red-sequence method \citep{gladders05}:  using the $z' - 3.6\mu$m color, which spans the 4000$\AA$-break at $z>$ 1, it locates over-densities of red-sequence galaxies that trace  massive galaxy clusters. This simultaneously provides a reliable redshift estimate for the cluster, through the color fit to the red-sequence, discussed in more detail in the Section 3.

SpARCS  uses the public SWIRE images and catalogs which provide photometry measurements at 3.6, 4.5, 5.8 and 8.0 $\mu$m from the IRAC instrument and 24$\mu$m with the MIPS camera (Londsdale et al.~2004; Shupe et al., private communication).  
SWIRE contains northern and southern sky coverage; however to ensure uniform ancillary data (described below) we limit this work to the 34 square degrees of the northern fields (ELAIS-N1/2, XMM-LSS, Lockman), and further require uniform coverage in IRAC 3.6$\mu$m and 4.5$\mu$m. 

\begin{figure}

\includegraphics[scale=0.5]{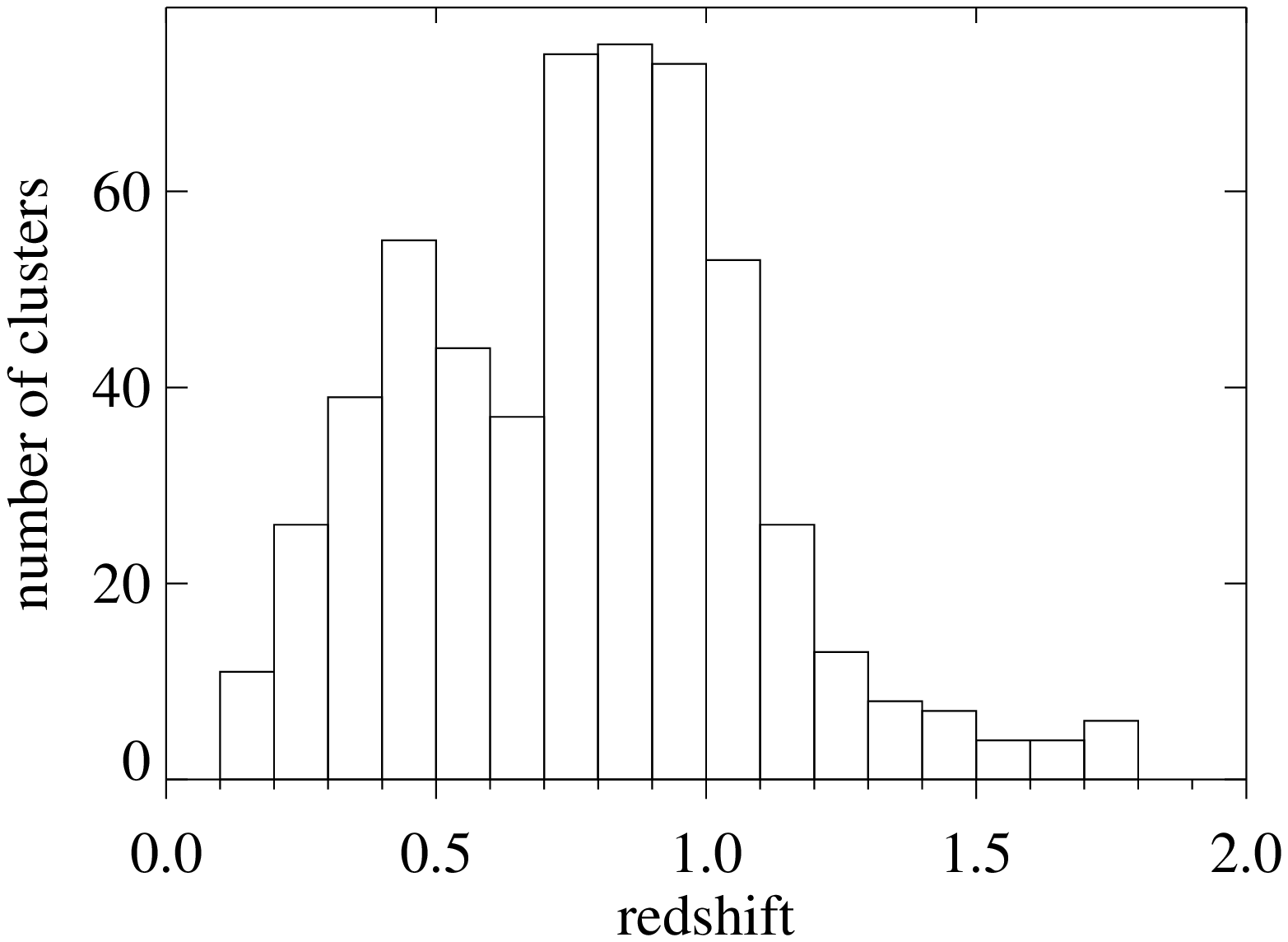}
\includegraphics[scale=0.5]{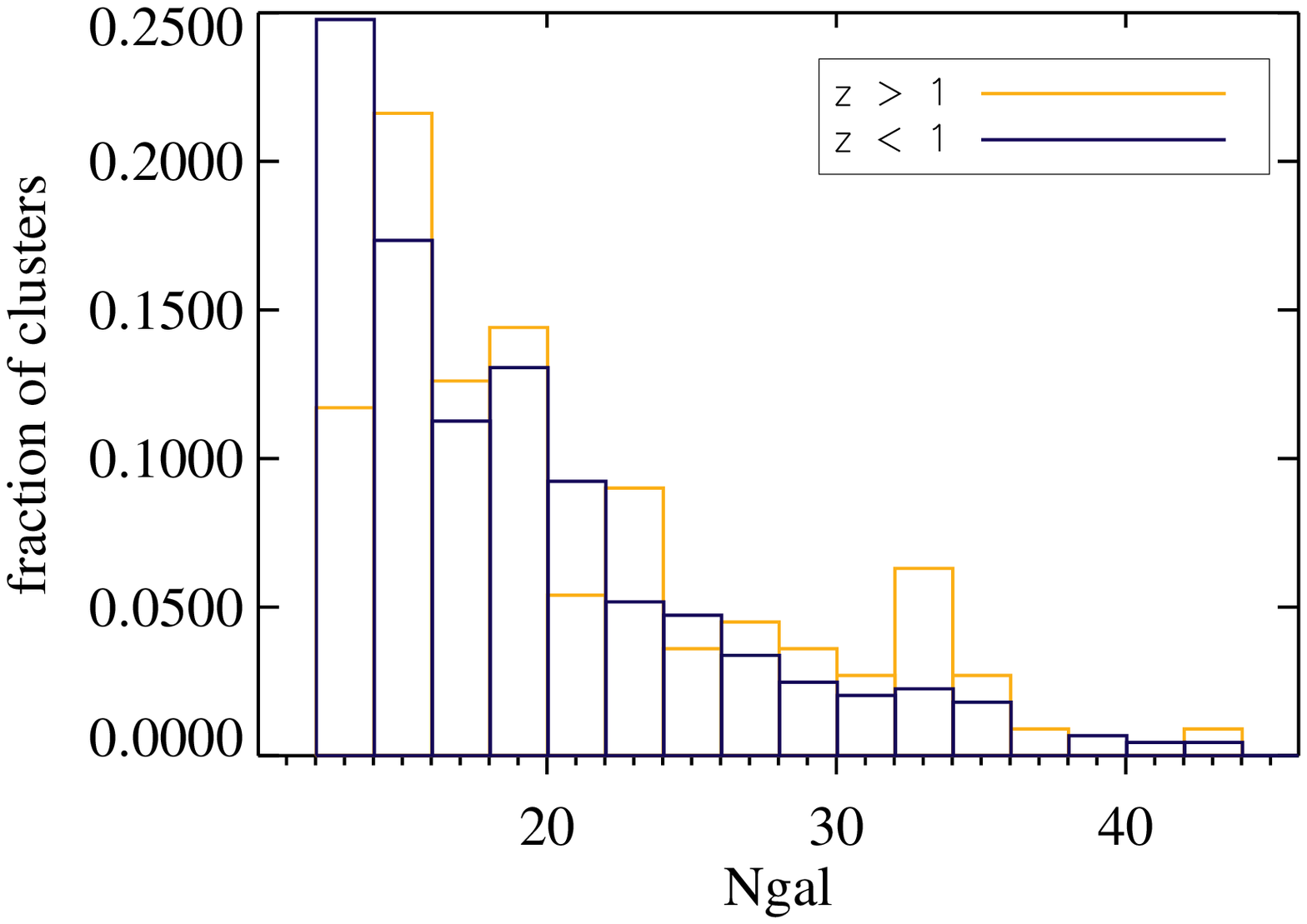}

\vskip 0.5cm
\caption{{\it Top:} The redshift distribution of the SpARCS cluster sample used here. {\it Bottomt:} The richness distribution (parameterized by N$_\mathrm{gal}$) of the $z< 1$ clusters (black) and the $z > 1$ clusters (orange).  Although fewer clusters are found at higher redshift, there is no significant difference in the richness distribution between the two redshift bins. \label{ngal}}
\end{figure}

 The parent sample is limited  by the detection reliability of the red-sequence over-density and the measured NIR richness.  Galaxy clusters must have a ``flux'' of at least 4 in over-density  in the detection map -- a criteria similar to that of  \citet{hildebrant11} --  and have a measured richness of $N_{gal} > $12. $N_{gal}$ is defined as the number of background subtracted galaxies within an aperture of 500kpc above a limit of $M_{3.6{\mu}m}^*$ +1.    These criteria ensure a richness-limited sample of high signal-to-noise detections (Figure \ref{ngal}).   To place the richness in  physical context we note that $N_{gal} = 12$  corresponds to $M_{200}$ $\sim$  1$\times$10$^{14}$$M_\odot$  \citep{wen10,wen12,capozzi12,andreon14}.  The sample of  559 galaxies is  further culled as discussed in Section 3.

\section{Analysis}
\subsection{BCG Identification}

Given the large number of clusters  (a final sample of 535, 106 confirmed and  429 candidates) within our sample, we automate the BCG identification algorithm.   We select the brightest galaxy in the IRAC 3.6$\mu$m channel whose color is  within $\pm$0.5mag of the  predicted red-sequence. The location red-sequence is defined by the $z'-3.6\mu$m color models of \citet{bruzual03} following \citet{muzzin09}.  Using clusters in common we check this against the smaller, but more detailed, study of high-redshift BCGs of \citep{lidman12}, and find  that they select the same objects.  This method is, however, susceptible to contamination by foreground dusty galaxies which are likely to be bright at 24$\mu$m and may have colors sufficiently reddened so as to be consistent with the red-sequence at higher redshift. Since the goal of this work is to study the infrared properties of the BCGs it is imperative that the frequency of such interlopers be quantified, and the
 misidentified BCGs removed from the analysis.  
 
 We identify candidate interlopers through visual inspection of the galaxy morphology. We use the Sloan Digital Sky Survey (SDSS) as it provides uniform imaging over all four fields. We identify  galaxies with obvious spiral structure and remove these from the analysis. We perform a secondary check using the the deeper imaging of the Canada-France-Hawaii Telescope Legacy Survey (CFHTLS)   in the XMM-LSS field, but do not identify any additional interlopers. In total, 24/559 galaxies are removed leaving a final sample of 535 BCGs.

\begin{figure}
\includegraphics[scale=0.5]{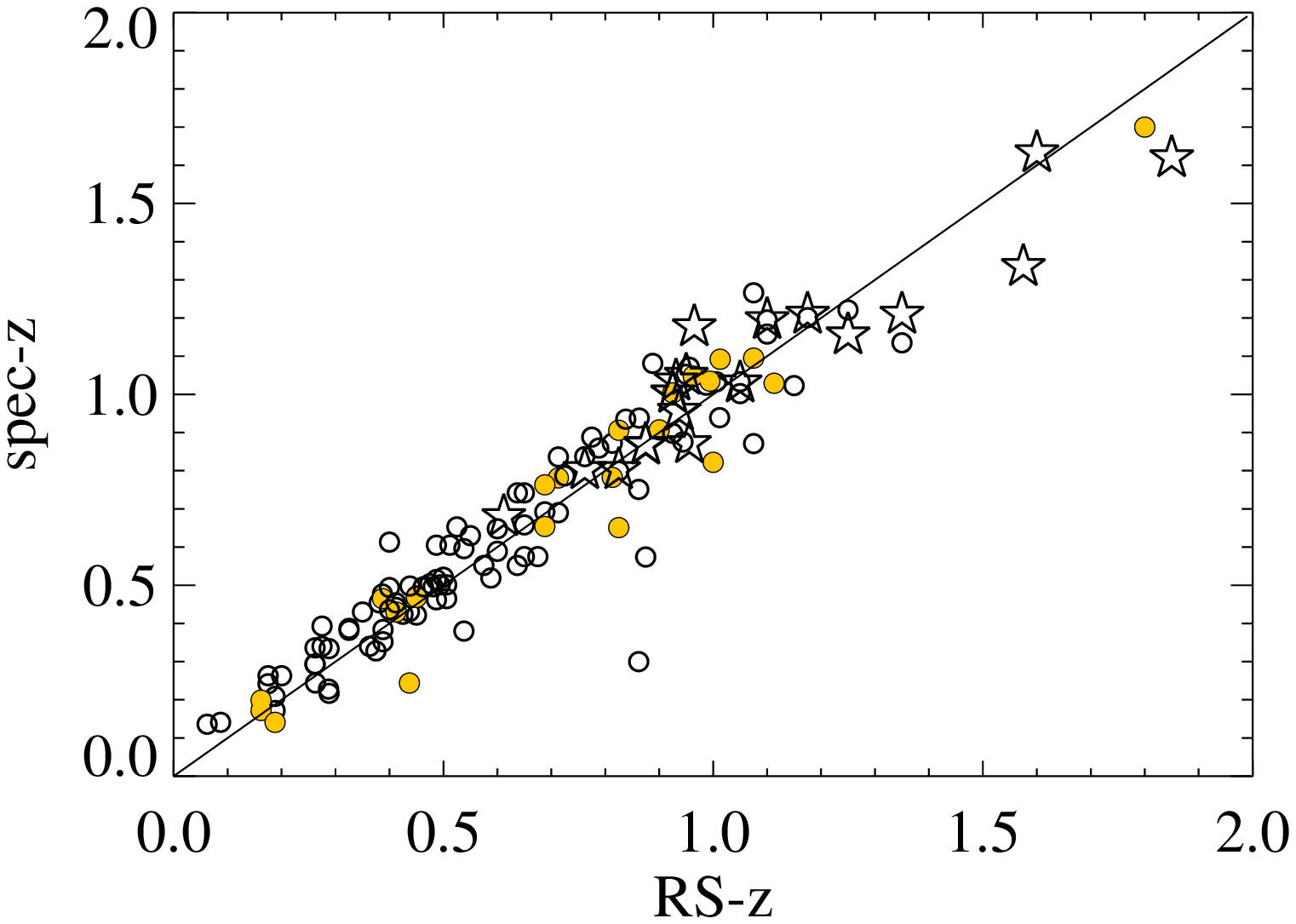}

\caption{A comparison of the cluster redshift, estimated through the location of the red-sequence, and the BCG  spectroscopic redshift (open circles) or cluster spectroscopic (open stars) redshift where available \citep{wenger00,rowan08,coil11,muzzin12,cool13,rowan13,fang15}.  The solid line denotes a 1:1 correspondence.  The BCG catalog has been cleaned by-eye to remove all obvious low-redshift galaxy mis-identifications. Solid orange points show BCGs with 24$\mu$m counterparts (Section 3.3), and these do not have a systematically different redshift offset than the  24$\mu$m non-detected population. The overall scatter in red sequence redshift is $\sigma$ = 0.10, using 106 spectroscopic redshifts. \label{redshifts}}
\end{figure}

 We  then compare the  spectroscopic redshift of the candidate BCG  when available with the red-sequence-estimated redshift of the cluster.  This comparison is shown in Figure \ref{redshifts} and includes spectroscopic 106 redshifts assembled from SDSS,  our own work \citep{muzzin12}, the OzDES survey \citep{fang15}, PRIMUS \citep{coil11,cool13},  and the literature \citep{rowan08,rowan13,wenger00}. A systematic offset of $z$(RS) - 0.15 has been applied to the entire sample. This offset was empirically determined from these 106 redshifts to correct for differences between the model and measured colours of the red sequence. We are, as of yet, unsure  of the source of this offset, but this will be further explored in Muzzin, A. et al., in prep.  As can be seen in Figure \ref{redshifts} there is excellent agreement between redshifts and no additional galaxies are removed based on redshift disagreements.  Although there are several BCGs whose RS redshift estimates are low compared to the overall  scatter within the population,  there is only one catastrophic outlier ($|\Delta z| > $ 0.5).   This small offset may reflects a tendency of the RS method to occasionally overestimate the redshift, rather than a misidentification of the BCG.  After all  interlopers are removed,
 we are left with a final sample of   535  BCGs.   The richness of their parent clusters span $N_{gal} = 12-40$ at all redshifts and the redshift distribution is shown in Figure \ref{ngal}.

  \subsection{Caveats on Identification of the Brightest Cluster Galaxy}

  This method of BCG identification has two additional biases.  Firstly, it is biased against BCGs with unusual colors - i.e. colors that are inconsistent with the red-sequence. Thus, if a BCG were exceptionally blue or red,  for example due to  star-formation (unobscured or obscured), it would be passed over by the algorithm, and instead the second brightest red-sequence galaxy in the cluster would be chosen (maintaining the redshift agreement but technically selecting the wrong galaxy).    In contrast to foreground contamination,  which adds spurious star forming galaxies to the sample, this effect may remove real star-forming BCGs from the sample.  
  
 Secondly, the sample could be contaminated by late-stage mergers that are unresolved  in the IRAC-3.6$\mu$m imaging (with a Point Response Function of 1.8$''$). Such systems, which are actually two galaxies, but still cluster members, may appear as the brightest galaxy within the cluster when their light is combined.  This latter effect would be more important at higher redshift where the apparent separation is smallest.   We see no signs of this in the optical imaging, but higher resolution observations of a significant fraction of the sample are required to confirm this.
  
Aside from the practical difficulties in identifying the BCG of a distant galaxy cluster, we must also consider the evolving definition of the BCG itself.  N-Body and SAMS \citep{delucia07} indicate that BCGs do not exist as a single identifiable  
object above $z\sim$ 0.7, and this is supported by some, but not all,  observations of distant clusters \citep[e.g.][]{santos09, webb15}. In these models, where the BCG assembles through dry mergers and galaxy cannibalism, 
the stellar mass which will eventually form the BCG is distributed among more than one galaxy.  Even so, these galaxies will  still be among the most massive systems in the cluster and our algorithm would select the brightest of them.   Thus, our approach may be more accurately described as identifying one of the progenitors of the  BCG galaxy.  An investigation of the uniqueness of BCG galaxies at high redshift is possible with these data, but beyond the scope of this paper and will be the focus of later work by our group.

\subsection{24$\mu$m-detected BCGs}

The BCG catalog was cross-referenced with the SWIRE IRAC-24$\mu$m imaging catalogs, with a MIPS-BCG search radius of 3$\arcsec$, to account for the uncertainty in the MIPS positions due to the 6$''$ Point-Response-Function. Table 1 lists the number of 24$\mu$m-detected BCGs in each field, to the depth of the SWIRE images ($\sim$ 150$\mu$Jy), with a total of  125 detected systems, or $\sim$23\% of the galaxies.   Table 2 provides the coordinates, redshifts, and 24$\mu$m flux measurements of these systems.

\subsubsection{Spurious Alignments}
We estimate the number of spurious spatial coincidences through simple Monte-Carlo simulations.  Using the 24$\mu$m catalog we repeat our BCG matching, but with random positions in place of the BCG locations. The number of expected spurious detections are listed for each field in Table 1 and indicate a spurious  BCG-24$\mu$m contamination rate of $\sim$ 1\% - much lower than our detection rate.  We therefore do not attempt to correct for it in further analysis.

\begin{deluxetable}{lccc}
\tablewidth{0pt}
\tablecaption{Information on cluster sample and observations \label{data_table}}
\tablehead{
\colhead{SWIRE Field}             & \colhead{Number } &
\colhead{Number } & \colhead{expected }    \\
\colhead{} & \colhead{of BCGs\tablenotemark{a}} & \colhead{ of BCGs } & \colhead{spurious}  \\
\colhead{ } & \colhead{in sample} & \colhead{with S$_{24{\mu}m}  > $ 100$\mu$Jy}     & \colhead{matches}  }

\startdata
XMM-LSS   &  124 &  29  &  2  \\
Lockman & 202  &  43   & 2 \\
ELAIS-N1   & 145  & 36 &  2 \\
ELAIS-N2   & 64 & 17  & 1 \\
\hline
total & 535 & 125 & 7 \\

 \enddata
\tablenotetext{a}{With the cluster richness N$_\mathrm{gal}>$ 12.} 
%make this footnote extra long so that it extends over two lines.}
%% You can append references to a table using the \tablerefs command.
%\tablerefs{
%(1) Barbuy, Spite, \& Spite 1985; (2) Bond 1980; (3) Carbon et al. 1987;}
\end{deluxetable}

\subsubsection{Physical Misidentifications: Gravitationally Lensed Systems and Close Neighbours}
We classify as physical misidentification cases where the IR emission is not associated with the BCG itself, but with an object which is physically related to the BCG.  Here we consider gravitational lensing and close neighbours. 

Strong gravitational lensing from the galaxy cluster potential and/or the central massive galaxy can magnify background IR galaxies along the line of sight to the cluster centers. 
  If the offset between the lensed object and the BCG is less than $\sim$3$''$  the IR flux will be confused within the Spitzer-MIPS  beam and erroneously assigned to the BCG through the matching procedure.  

\begin{figure}
\includegraphics[scale=0.5]{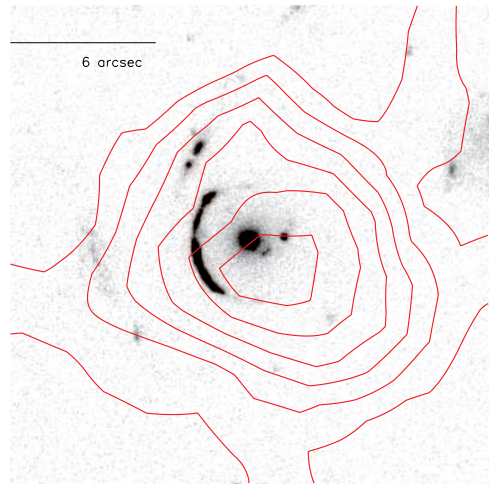}

\caption{The HST-ACS-606W image one of the BCGs in this sample (from the CANDELS UDS public data) that is clearly lensing a background galaxy.  In red we show the MIPS flux contours starting at 0.05 MJy/sr and increasing in 0.05 steps. Given the beam size of MIPS (6$''$ FWHM shown) the source of the 24$\mu$m emission is ambiguous. \label{hst}}
\end{figure}

To search for gravitational lenses within the SpARCS-BCG sample we turn to archival or public HST imaging. Very few  (10) of the sample presented here have been covered in past surveys, but that number increases if we include the less rich ( but still meeting the significance criterion of Section 2.1) systems in SpARCS.   In doing so we find one example of a tight ($\sim$2$\arcsec$) lensing system in one cluster (Figure \ref{hst}).   If this rate ($\sim$10\%) is representative of the entire sample, lensing  would indeed be an important source of contamination.   However, this object is a previously known  system that was first discovered as a   strong galaxy-galaxy lens by \citet{geach07} and later imaged in  UDS field of the CANDELS survey \citep{grogin11,koekemoer11}.  Its HST coverage may  not be by chance, and therefore this rate may not be representative of the lensing rate.

To gauge if this lensing rate is at all reasonable, given what we know about the systems, we ran lensing simulations following the method of \citet{hezaveh12}.   We find the  small radius lensing rates  of only $\sim$ 1\%, which are too low to account for the IR detection rates of  23\%, and moreover, the rate does not increase with redshift.   While the lensing rates can be fine-tuned by altering the simulations (for example by tuning the ellipticity or BCG-cluster alignments), we conclude that lensing is not a major contaminant of our sample.

A second  possible source of contamination that is physically interesting is the presence of close neighbours to the BCG.  By this we refer to physical neighbours, rather than line-of-sight associations which would be accounted for in our Monte Carlo simulations of spurious detections.    These galaxies need to be close enough to be confused within a single MIPS 24$\mu$m PRF, but separated into two objects in IRAC 3.6$\mu$m.  At such close separation the objects will likely be interacting with the BCGs and thus since the two galaxies in a sense form a single system that includes the BCG, they modify but do not negate the interpretation of  our results.  

To assess this issue, we have compared  the frequency of close galaxy neighbours in the  24$\mu$m-detected and undetected samples using the full IRAC SWIRE galaxy catalog and see no difference between the two, however  a proper assessment of this effect requires higher resolution imaging so that very close neighbours can be revealed.
\section{Results}

\subsection{MIR Diagnostics of the Energy Source: AGN versus Star Formation}

\begin{figure}
\includegraphics[scale=0.5]{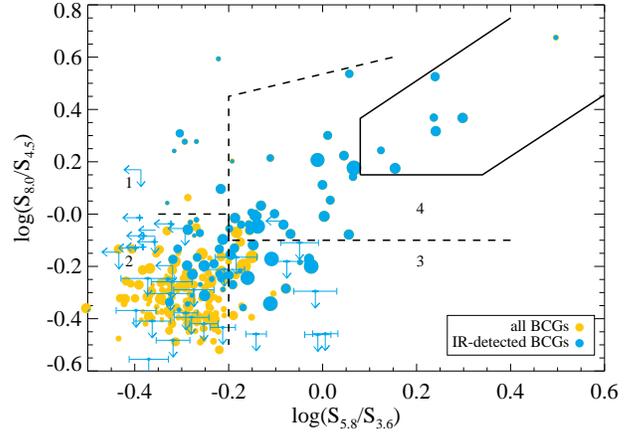}
\caption{The IRAC color-color plot for the SpARCS BCGs. Blue points correspond to 24$\mu$m detected BCGs, and orange to 24$\mu$m non-detected BCGs.   The increasing sizes of the  points corresponds to increasing redshift. We also show 24$\mu$m detected BCGs that have limits on their 5.8$\mu$m or 8.0$\mu$m flux, but for clarity only include the infrared-faint BCGs that have detections in all four IRAC channels. Overlaid are the \citet{sajina05} SED template sections (dotted divisions; 1: low-$z$ PAH; 2  and 3: mid/high-$z$ PAH and stellar continuum; 4: high-$z$ PAH and all-$z$ AGN) and the more recent \citet{donley12} AGN wedge (solid line). \label{zoom}}
\end{figure}

 In Figure \ref{zoom} we show the 4-channel IRAC colors for the 24$\mu$m-detected and the 24$\mu$m- undetected  BCGs.
The IRAC colour plot is frequently invoked as a tool to discriminate between AGN (continuum-dominated),  star-forming (PAH-dominated) and passive (stellar-dominated)  galaxy systems \citep{sajina05, lacy04, lacy07, stern05, donley12}.   In Figure 4 we show the divisions suggested by \citet{sajina05} (1: low-$z$ PAH; 2  and 3: mid/high-$z$ PAH and stellar continuum; 4: high-$z$ PAH and all-$z$ AGN) to statistically separate these populations as well as the more recent, and tighter, AGN selection criteria of \citet{donley12}.  

The IR-faint BCGs lie almost entirely within regions 2 and 3, indicating that they are truly passive systems, or higher redshift star forming galaxies with star formation rates too low to be detected in SWIRE.  The low redshift star-forming  region is relatively empty, containing only one IR- undetected and seven IR-detected systems.  The cluster redshifts for these systems  are  $z<$ 0.6, in agreement with the expected low redshifts.  The bulk ($\sim$75\%) of the IR-detected objects lie within regions 1, 2 and 3 with higher redshift systems preferring region 3; in  agreement with the interpretation that their IR emission is star-formation dominated.  Region 4 contains only (bar one object) IR- detected systems. In this region we find a combination of high-redshift star-formers and AGN. Indeed, many of these systems form an extension of the colors of regions 2 and 3, suggesting they are also dominated by starbursts.  

 Subsequent analysis  identify as AGN only those 7 systems which lie within the updated  constraints of \citet{donley12},  but otherwise include BCGs in region 4.  An upper-limit  to the contamination from AGN to the MIPS detections can be obtained by considering all galaxies in region 4 as AGN.

\subsection{Mid-Infrared Luminosities of  BCGs}

We convert the observed 24$\mu$m fluxes to inferred total IR luminosities following the models of \citet{chary01}.  
This assumes that all of the 24$\mu$m flux is emitted by the BCG itself (that is, no gravitational lensing) and is entirely due to star formation with no AGN contamination.  We therefore highlight, when relevant,  those galaxies which may contain AGN, as identified in the previous section (Donley et al.~2012)

\begin{figure}
\includegraphics[scale=0.5]{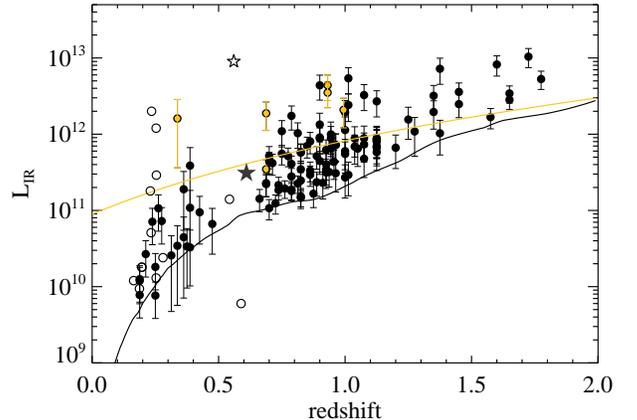}

\caption{The 24$\mu$m-inferred infrared luminosity (assuming the models of \citet{chary01}.)  Solid points refer to the SpARCS cluster sample (this work), with the orange colour denoting possible AGN, as identified by the Donley et al.(2012)  region in Figure \ref{zoom}.  The error bars include the 24$\mu$m flux uncertainties and a $\pm$0.1 redshift scatter. The open circles correspond to the Herschel detected BCGs of \citet{rawle12}, although we use their 24$\mu$m estimated $L_{IR}$ for consistency.  The open star denotes the extreme star-forming BCG reported by \citep{mcdonald12}, discovered in  the SPT survey.  The solid star corresponds to the BCG with a strongly lensed arc (Section 3.3.2); here we assume the source of the 24$\mu$m flux is the BCG itself. Note that this object is not actually part of our study as it has $N_{gal} < $ 12, but is shown for reference. The orange solid line shows the luminosity depth discussed in Section 4.2, that is tied to an evolving IR luminosity function, and the solid black line the approximate luminosity depth of the SWIRE 24$\mu$m survey.  \label{fir}  }
\end{figure}

In Figure \ref{fir} we show the infrared luminosity of the 24$\mu$m-detected BCGs with redshift  and we list the values in Table 2. We include for reference the  Herschel detected BCGs of \citet{rawle12} and  the extreme star forming BCG in the Phoenix cluster reported by \citet{mcdonald12}.   The Phoenix  BCG is an outlier at $z=0.6$, whereas the IR luminosities of the Rawle et al.~sample are comparable to ours, with a few exceptions.  Note however, that the Rawle et al.~clusters are X-ray selected and located at  redshift $z < 1.0 $, whereas our sample is effectively stellar-mass selected to  higher redshift, and therefore a direct comparison of the two samples is not straightforward.

\begin{figure}
\includegraphics[scale=0.5]{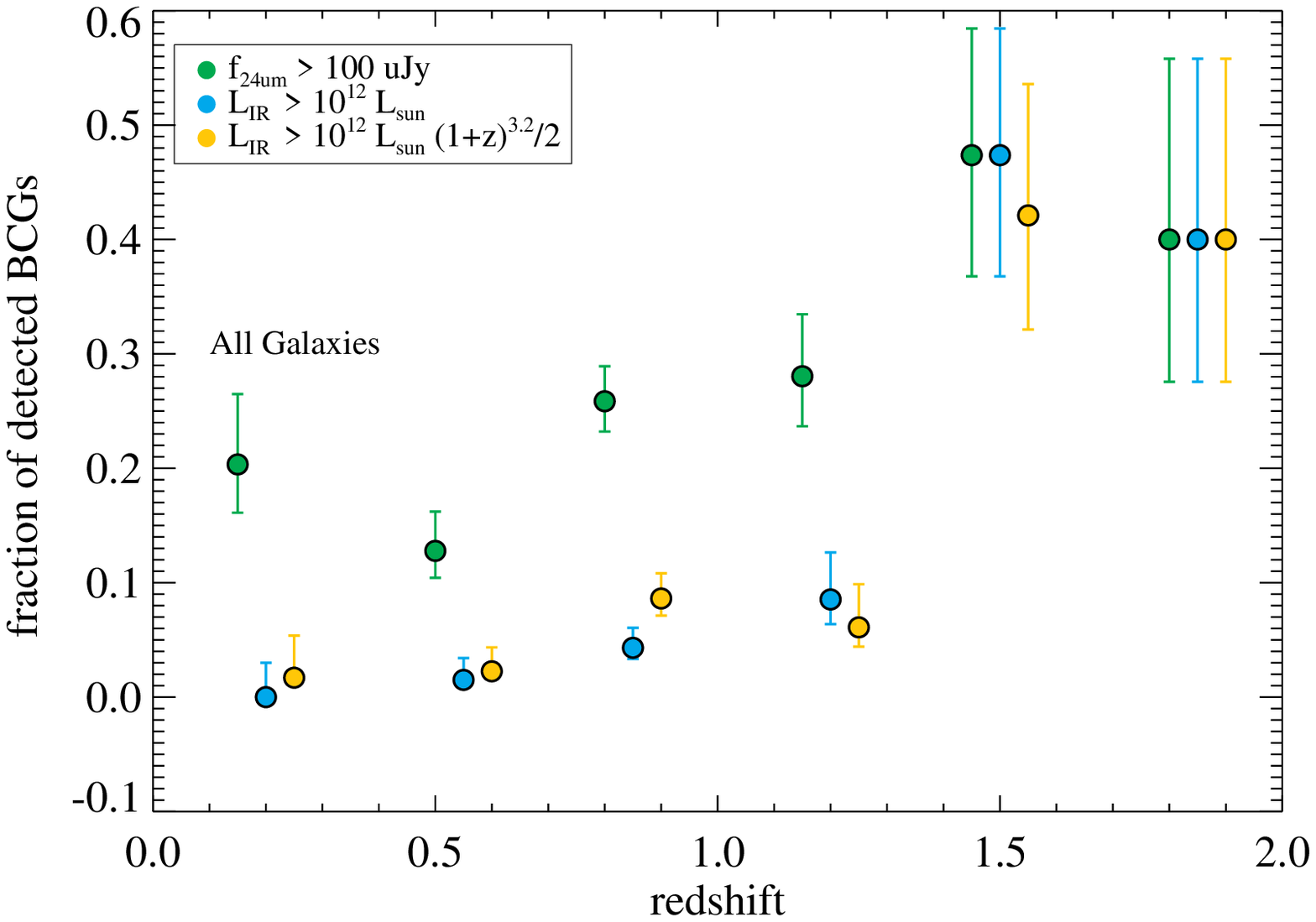}
\includegraphics[scale=0.5]{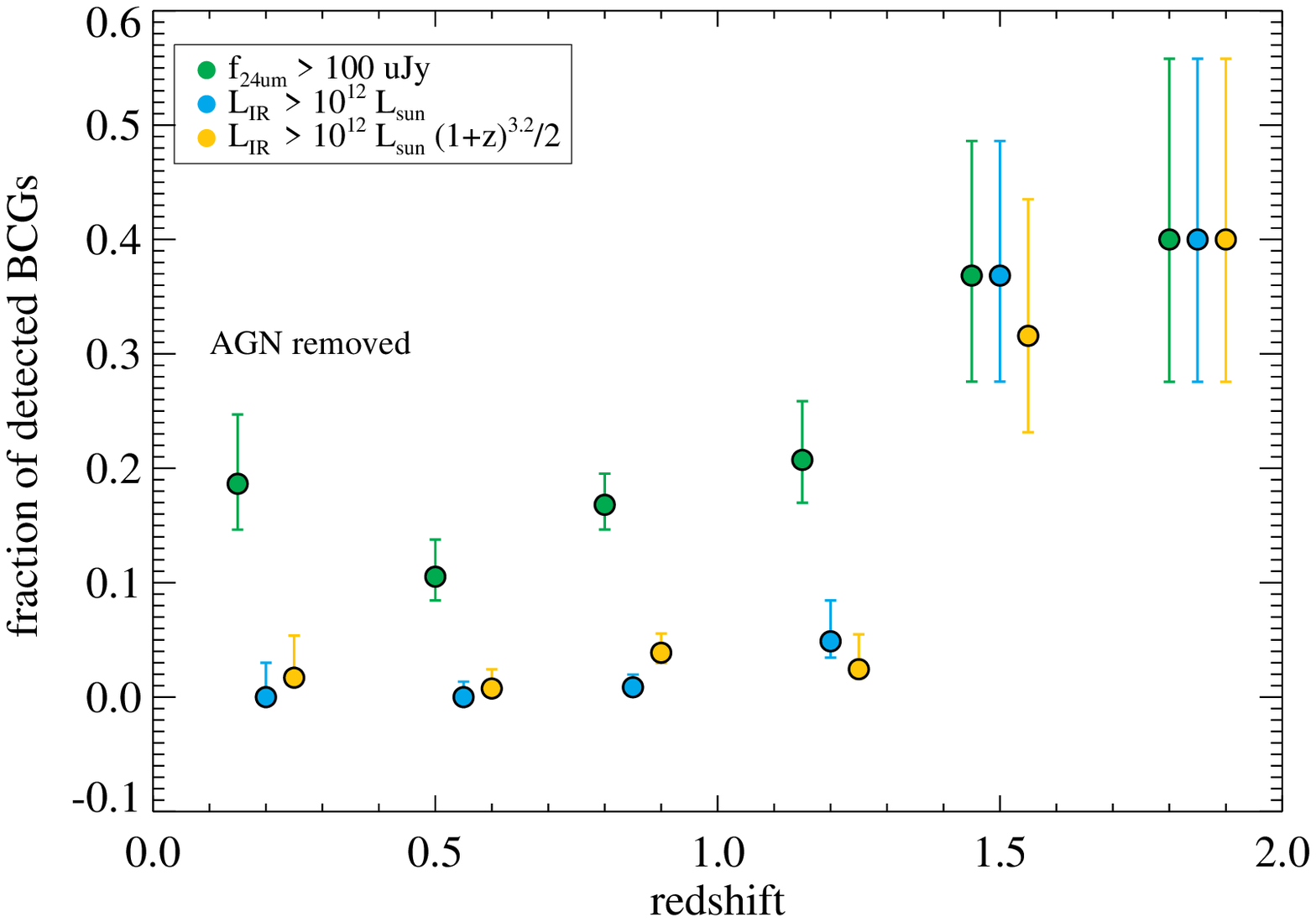}

\caption{{\it Top:} The fraction of 24$\mu$m detected BCGs in different redshift bins.  Blue points denote BCGs with inferred infrared luminosities of $>$$L_{IR}$ 10$^{12}$ $L_\odot$ and orange points denote the fraction of BCGs above the evolving luminosity cut shown by the orange line in Figure \ref{fir}.  Also shown (green) are the fractions of BCGs detected at any 24$\mu$m flux above the detection limit of the SWIRE survey.   {\it Bottom:} Same as above, but with possible AGN contaminated galaxies removed from the analysis. For this plot we define as AGN those galaxies lying in region 4 of Figure 4. \label{frac} }
\end{figure}

%\begin{figure}
%\includegraphics[scale=0.5]{frac_bin_agn.ps}
%\caption{The same as Figure \ref{frac},.   \label{frac_agn} }
%\end{figure}

There is a consistent rise in Figure \ref{fir} in the inferred infrared luminosities of BCGs with similar cluster richness out to $z\sim$ 2.  Given that the infrared luminosity function of field  galaxies evolves toward higher luminosities and densities  over this redshift range, this evolution may simply be a reflection of this general trend.   In Figure \ref{fir} we show an evolving infrared luminosity limit  following the $L_\mathrm{IR} \sim $(1+z)$^{3.2}$ relation of Le Floc'h et al. (2005) and scaled to our highest luminosity  depth ($\sim$$L_{IR}$ = 1$\times$10$^{12}$ $L_\odot$ at $z=1.8$). This defines the luminosity limit as the same fraction of an evolving $L_{IR}^*$, at every redshift, where, for reference $L_{IR}^* \sim$ 1.6$\times$10$^{10}$ $L_\odot$ at $z\sim$ 1.

 There is still an increase in the absolute number of objects above this line, however this should not be interpreted as a luminosity evolution;  the absolute numbers of BCGs at a given luminosity will depend on the volume probed (which increases by $\sim$ 3 from $0 < z< 1$ to $1 < z < 2$), as well as the cluster number density and selection function.

In Figure \ref{frac} we show the  fraction of BCGs that are detected above a given luminosity limit which is more likely to trace a real change in clusters with time.  In the top panel of Figure \ref{frac} we show  the fraction of all  BCGs detected at different luminosity/flux  limits.  We include the fraction of BCGs detected at any flux above the depth of SWIRE ($\sim$100$\mu$Jy);    above $L_\mathrm{IR} > $ 10$^{12}$ $L_\odot$; and above the field related evolving infrared limit shown in Figure \ref{fir}.      In all cases the fraction of IR bright  BCGs  increases significantly beyond $z\sim$1, for similar richness clusters.  In particular, the field-corrected evolution increases from $\sim$ 5\% to  $\sim$ 30\%, from the lowest to highest redshift bins.  This plot includes all IR detected systems, even those with possible AGN contamination.  In the lower panel we show the trend after removing all galaxies within the Sajina et al. (2005) region 4,  which as explained in Section 4, we take to be the highest level of AGN contamination.  Using the Donley et al. region provides a much lower   contamination rate.   While the evolution is reduced by roughly a factor of two (for the Sajina et al. contamination rate), it remains significant, in part because the AGN are scattered throughout most of the  redshift bins.

We note, finally, a systematic effect that may be present in this analysis. Several studies have shown that the \citet{chary01} methodology over-estimates the infrared luminosity above a redshift of $z\sim$ 1.5 \citep{murphy09,nordon12,rodighiero10} by roughly a factor of 5, due to large PAH equivalent widths at these redshifts. This redshift transition is the approximate location where we see the largest change in the fraction of 24$\mu$m-detected BCGs. It may be that, due to this effect, we are sensitive to a  lower luminosity limit beyond $z\sim$1.5, which would in turn lead to a larger fraction of detected galaxies.  If we applied a downward correction to the $L_{IR}$ in Figure 5, this would indeed flatten the high-redshift tail and bring those estimations more in-line with the measurements below $z\sim$1.5, but it would not change our results qualitatively as these systems would still remain at $L_{IR}>$10$^{12}$$L_\odot$. Moreover, the infrared luminosity of  the most luminous object in our sample, SpARCS1049+56 at $z=1.7$, has been studied in detail in Webb et al. (2015), where we determine its luminosity using  6 infrared measurements and two limits. We determine the $L_{IR}$ to be 6.6$\pm0.9 \times$10$^{12}$ $L_\odot$, compared to an $L_{IR}$ to be 1.0$\pm 0.3 \times$10$^{13}$ $L_\odot$ determined from the 24$\mu$m measurement alone.  These estimates are within 1$\sigma$ of each other indicating, at least for this particular object, there is no strong over-estimate of the flux.

\subsection{Star Formation Rates of the BCGs}

\begin{figure}
\includegraphics[scale=0.5]{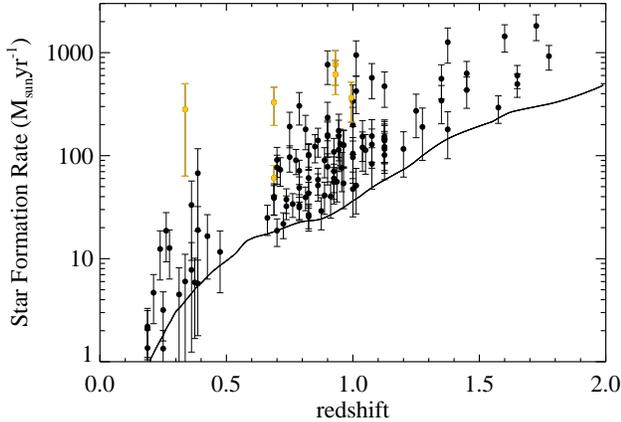}
\caption{The inferred star formation rates of the BCGs with redshift.  The orange points correspond to those  with IRAC colours consistent with AGN following the \citet{donley12} criteria. The solid line shows the approximate depth of the SWIRE 24$\mu$m imaging.  \label{sfr}}
\end{figure}

We can scale  the infrared luminosities determined in the previous section to star formation rates following the relation of \citet{kennicutt98}. Based on Figure \ref{zoom} we assume no AGN contribution to the infrared flux, which would systematically reduce the estimates.   We show the SFRs for each BCG with redshift in Figure \ref{sfr}.  Uncertainties include the photometric uncertainty in the MIPS 24$\mu$m flux and the redshift uncertainty  of $\Delta z\pm$0.1. A wide range in star formation rates is seen, from $\sim$1 at low redshift to  $\sim$1000 $M_\odot$yr$^{-1}$ for the highest redshift BCGs.  

 Using stellar mass measurements for the BCGs we can further convert the star formation rates to star formation rates per unit stellar mass, or specific star formation rates (sSFRs).   Given the uniformity of the BCG optical colours we adopt a simple methodology to measure stellar mass.  We use the observed 3.6$\mu$m flux to determine the rest-frame $K$-band luminosity, adopting an 11 Gyr  single stellar population   from \citet{bruzual03} to calculate the $K$-correction. We then take the average $K$-band mass-to-light ratio of a red galaxy (M/L$_K$ = 1) from \citet{bell03}, with a scatter of 0.1dex. To determine the sSFR we further scale the masses by 1.65 to convert to a Salpeter IMF \citep{raue12}.  

\begin{figure}
\includegraphics[scale=0.5]{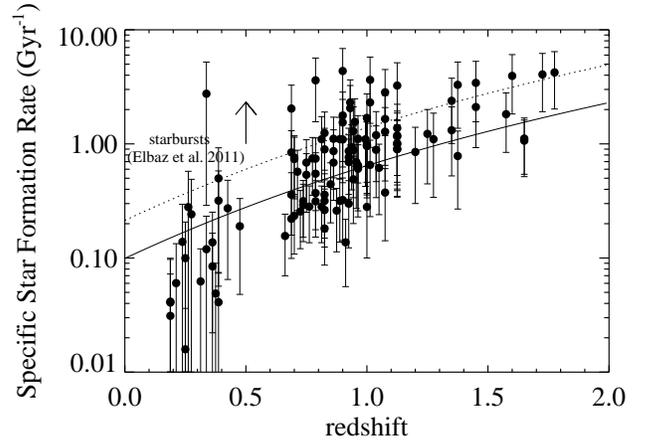}
\caption{ The specific star formation rate  of the 24$\mu$m-detected BCGs shown as a function of redshift. Overlaid (solid line) is the best fit relation of \citet{elbaz11} for main sequence star forming galaxies. The dotted line shows the rough division between starburst  galaxies and main sequence galaxies, as classified by Elbaz et al.   No AGN correction has been made to the SFRs.\label{ssfrz}  }
\end{figure}

Figure \ref{ssfrz} shows the sSFRs of the individual BCGs with redshift.    This figure also contains the evolution of the sSFR of the main sequence of star forming galaxies \citep{elbaz11} (solid line), as well as the division between main sequence and starburst systems as defined by Elbaz et al. 

%{  Uncertainties include the photometric uncertainty in the MIPS 24$\mu$m flux, the redshift uncertainty  ($z\pm$0.1)  , and a 0.1 dex uncertainty in the stellar masses \citep{bell03}. The solid line corresponds to  the evolving location of main sequence star forming galaxies from \citet{elbaz11}, and the dotted line to the approximate division between main sequence galaxies and starbursts, as classified by Elbaz et al.   At all redshifts the sSFRs of the BCGs are consistent with main sequence galaxies. That is,  although the individual star formation rates are high, the correspondingly high stellar mass of the systems, and the overall evolution of the field to higher sSFRs, results in main sequence level sSFRs at each redshift.  }

\begin{figure}
\includegraphics[scale=0.5]{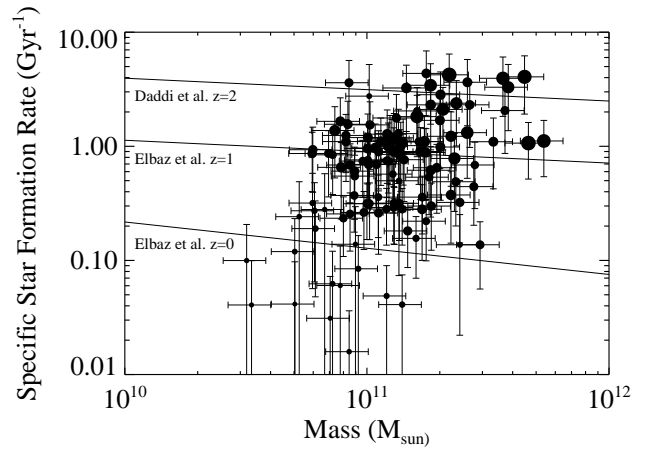}
\caption{ The specific star formation rate of the 24$\mu$m-detected BCGs shown as a function of stellar mass. Overlaid are the relations for main sequence star forming galaxies at the redshifts $z = 0,1$ \citep{elbaz07} and $z =2$ \citep{daddi07} that encompass our sample.  The size of the data point, for each BCG, increases with increasing redshift, from $z \sim$0.1-1.8.   \label{ssfrmass}  }
\end{figure}

  In Figure \ref{ssfrmass} we show the sSFR of the 24$\mu$m-detected BCGs as a function of stellar mass.   The size of the points increases with the redshift of the BCG so that the largest points correspond to  at $z = 1.8$ and the smallest points to $z = 0.1$. Overlaid we show the location of the main sequence of star forming galaxies in the field for the three redshift regimes that encompass our sample:  $z = 0, 1, 2$ \citep{elbaz07,daddi07}, again corrected to a Salpeter IMF. 
 
In both Figures \ref{ssfrz} and \ref{ssfrmass}  general agreement is seen between the sSFRs of the BCGs and the overall level seen in the field at a given redshift or mass.  This indicates that although the star formation activity within the BCGs is  quantitatively high, it is not in great excess of that seen for  field galaxies,  and indeed the BCGs may be classified as main sequence star forming systems. 


    \section{Discussion}

\subsection{A Change in the Observed Activity  in BCGs with Redshift}

These data show a measurable  change in the IR properties of  similarly selected BCGs from redshift $z\sim$ 0.3 to $z\sim$ 1.8. 
The IR colors indicate that the bulk of this evolution  is due to an increase in the star-formation rates of BCGs to higher redshift, and the remainder from dusty AGN.  This is qualitatively similar to the rapid increase of  the global star formation rate density of the universe to $z\sim$ 1, as well as the migration of $L_{IR}^\star$ to higher luminosities over the same redshift range. Figure \ref{frac} suggests, however, that the BCGs may follow a steeper evolution than seen for the field.     We cannot constrain the slope of this evolution, because we are only sensitive to the most luminous galaxies at high redshifts, however we do see clear increase in the fraction of BCGs detected above an IR-depth which is defined relative to the evolution of the field IR luminosity function (see Section 4.2).  Note that this result is not driven by the increase in AGN contamination:  it remains when all of the candidate AGN are removed from the sample. 
This is  evidence that the physics driving the increased activity in BCGs at higher redshift is either different than that driving the field evolution or is accelerated in the cores of galaxy clusters. 

\subsection{The Stellar Mass Growth of BCGs}

Below $z \sim$ 1 the SFRs of individual detected BCGs do not exceed 100 $M_\odot$yr$^{-1}$.   If the detection rate of BCGs at this level ($\sim$10\%) is reflective of the duty cycle of the star-burst then they must be short-lived, or $\lesssim$ 1 Gyr.  At such rates, this star-formation will contribute a  relatively small amount of stellar mass to the overall system.  There have been conflicting measurements of the actual stellar mass growth rate of BCGs in this redshift range, but generally an increase between $\sim$ 0-2$\times$ is seen since $z = $ 1 \citep{lidman12,liu12}.   This mass growth has been attributed by many to the accretion of established stellar populations through gas-poor major or minor galaxy mergers \citep{edwards12,lidman13}.  The IR results seen here (below $z < 1 $) do not add enough stellar mass to contradict this scenario.

By $z\sim$ 1, however,  the star formation level has drastically increased to $\sim$ 500-2000 M$_\odot$/yr.  Using similar arguments as above we reach similar timescale conclusions:  if the duty-cycle is 20\% over the epoch studied, then the bursts must be limited to a few hundred million years.  In this case however, the greatly increased star formation rates mean that even over this short timescale a BCG can easily double its stellar mass.  At these levels in situ star formation is an important, perhaps dominant, contributor to the mass growth of BCGs \citep{collins09,stott11,lidman13}.

\begin{figure}
\includegraphics[scale=0.5]{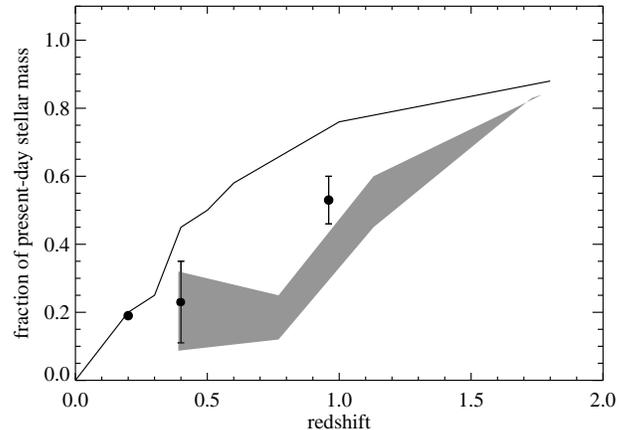}
\caption{This plot, which is adapted from \citet{lidman12},  shows the amount of stellar mass that is  added to BCGs with time, normalized to their present-day ($z=0$) stellar mass.  Solid  points denote the stellar mass measurements of BCGs from Lidman et al., and the solid line denotes the model predictions of \citet{delucia07}.  This is a modification of the Lidman et al.~plot;  here a value of 0.2 indicates that a BCG has a deficit of 20\% in its stellar mass, compared to today. 
The solid grey region shows the amount of stellar mass which, given the assumptions outlined in the text, would be added to BCGs by the dusty enshrouded star formation seen in this work.  The upper boundary is calculated by assuming all of the IR-bright systems detected here are star forming, while for the lower bound we have removed all possible AGN (Fig. \ref{zoom}) from the analysis.  
The intention of this plot is to compare the approximate mass growth through star formation with that required by the directly observed mass change (Lidman) and that expected from the models (De Lucia \& Blaizot). \label{mass_change}}
\end{figure}

This  general reasoning can also be  illustrated in Figure \ref{mass_change}, which is adapted from \citet{lidman12}.   Using NIR-determined mass measurements of BCGs Lidman et al.~calculated the fraction of $z = 0$ stellar mass in BCGs   in place at a given redshift, to $z\sim$ 1. They find relatively good agreement with the  build up of mass due to dry-mergers in the SAMS of \citet{delucia07}.    These two studies are shown in Figure \ref{mass_change},  but are plotted as the amount of mass still to be added to a BCG at a given redshift, as a fraction of the final mass.    We compare to this the  amount of stellar mass added by the star formation seen here, again in terms of the final stellar mass at $z = 0$ 
which is shown by the grey area. 

To determine the grey area we measure the  average SFR for the entire BCG population in redshift bins, assigning a SFR  of 0 $M_\odot$yr$^{-1}$ to the undetected BCGs. The chosen bins are $z=$0.2-0.6, 0.6-1.0, 1.0-1.4, 1.4-1.8. We then integrate this star formation over the duration of the redshift bin to obtain the average  amount of  stellar mass added in that bin. This is then combined with  existing average stellar mass of the BCGs within the redshift bin to provide a total, final mass, treating each bin independently.     This is complementary to our reasoning at the very beginning of this section where  we took the high star-formation-rates of the IR-detected BCGs and used the detected fraction to constrain the limits of integration (the duty cycle).  Here we do not constrain the timescale of the star-formation  -- allowing the star formation to proceed at the same rate over the entire redshift bin --, but employ a lower average star formation rate which incorporates the IR-faint galaxies as well.    The upper bound of the area includes all IR galaxies while the lower bound was computed with the candidate AGN removed.

Figure 10 is meant only to illustrate the approximate importance of the dust enshrouded star formation seen here, relative to the amount of mass required by other observations. The stellar mass determination is crude and no attempt is made to correct for progenitor bias, as done in Lidman et al.  Nevertheless, a simple picture emerges. 
 Out to $z\sim$ 1 in situ star formation  adds less to the total stellar mass of the BCGs  (as a population) than do the dry-merger predictions, and is less than required by the mass measurements of Lidman et al.     Above $z\sim$ 1 however, the situation is less clear. The  amount of mass added by star formation rises steeply and begins to approach  the mass assembled through dry mergers in the models.  At $z\sim$ 1.5 both processes would more than double the existing mass of the BCG.   More mass measurements of BCGs above $z\sim$ 1 are required to determine if there is actual tension between these two models of mass assembly.  Clearly both processes could operate and play significant roles in the formation of BCGs; however their combined effect is constrained by the observed buildup of stellar mass.

%These data employ the average SFR of all BCGs in redshift bins; the undetected BCGs are assigned a SFR of 0 M$_\odot$yr$^{-1}$.  This average SFR is 

\subsection{Comparison to Recent Literature}

The results presented here stem from the first infrared  analysis  of a large sample of BCGs at  $0.2 < z < 1.8$ They are, however, in  solid agreement with complementary studies of other massive galaxies and single clusters.  \citet{marchesini14} undertook a study of the progenitors of local Ultra Massive Galaxies (UMGs) within the UltraVista Survey using abundance matching techniques.   While not explicitly identified with clusters, the stellar-mass of 10$^{11.8}$ $M_\odot$ is similar to the mass of our BCGs   3$\times$10$^{10}$ - 7$\times$10$^{11}$, and it is likely the Marchesini et al.~population overlaps significantly with the population studied here.   They find remarkably consistent results:  the progenitors of UMGs remain quiescent until $z\sim$1, beyond which the exhibit extreme levels of dusty star formation of several hundreds of solar masses per year. 

Extreme star formation activity has also been measured toward or in the cores of several high-redshift clusters.  \citet{santos15} reports a total SFR of 1875$M_\odot$yr$^{-1}$ within the central 250kpc of a $z=$1.58 cluster, distributed over three massive galaxies.   Indeed, many groups are now finding  the star formation activity within the central regions of  $z>$1.5 clusters, though not always the BCG \citep{tran10,popesso12,fassbender11, fassbender14,brodwin13}, is consistent with the level of field activity. In some cases \citep[e.g.][]{fassbender14} this activity is due to an enhancement in galaxy-galaxy mergers within the cluster cores.

\subsection{Star Formation Driver:  Dissipative Mergers?}

These results of the previous sections indicate a rapid increase in star formation in the central galaxies of  clusters beyond $z \sim$ 1 and this in turn suggests a increase in the efficiency of gas deposition onto these systems.   In the field,  major galaxy mergers appear to be the dominant method of delivering new (and large) reservoirs of gas to galaxies, but this process may be problematic in very high density regions.   Galaxies within the cluster environment are subject to ISM removal processes such as ram pressure stripping or strangulation and may be  gas deficient \citep[e.g.][]{gavazzi06} by the time they reach the centre.   As pointed out in \citet{mcdonald12} for the  Phoenix Cluster BCG,  the prodigious star formation rates measured here require the  accretion galaxy with a gas mass far in excess (10-100$\times$) that of a typical ($z=0$) late-type cluster galaxy.
Certainly in-falling field galaxies at high redshift will begin their accretion with  more gas than at low redshift \citep{geach11}  and therefore may be able to retain a larger overall mass of gas during their infall. Indeed,  simulations of galaxy-ICM interaction indicate  galaxies may retain substantial fractions of their  gas during infall to the cluster center \citep{jachym07}, but the gas content of cluster galaxies beyond $z\sim$ 1 remains unconstrained.

Still, the one high-redshift and 24$\mu$m-bright for which we have conducted high spatial resolution imaging with the HST shows clear indications of a merger \citep{webb15}.  This system, SpARCS1049+56, consists of a long (66kpc) tidal tail interspersed with clumps along its length.  It appears to originate from within the stellar halo of the BCG and exhibits copious amounts of star formation ($\sim$1100 $M_\odot$yr$^{-1}$).  This gas-rich merger would not have been apparent without HST NIR imaging and thus ruling out this scenario for the rest of the sample is premature.  If SpARCS1049+56 is indeed representative of the $z>$1 24$\mu$m-bright SpARCS sample, then wet gas-rich mergers can occur at the centres of clusters and indeed, are an important process in building the stellar mass of BCG at early times.

 On the other hand, studies of  cool-core clusters \citep{rawle12,liu12} reveal a strong correlation between the star formation rates of BCGs (limited to below $z\sim$ 0.5)  and their cooling times (as measured in the X-ray). The sample presented here is optical- richness selected  (nominally stellar-mass-selected) and  there is  no information on the  state of the gas in these systems.  X-ray studies of the Red-Sequence Cluster Survey \citep{hicks13}, which are limited to $z<$ 1, indicate that perhaps 10\% of optically selected clusters harbour cool-cores.  This fraction is enough to account for all IR-detected galaxies in our sample to $z\sim$ 1, but does require the number of cool-core clusters to increase substantially at higher redshifts. A detailed comparison of SpARCS optical and X-ray selected clusters has not yet been made at these redshifts. One SpARCS field (XMM-LSS) has X-ray coverage and we have compared the clusters found by the SpARCS survey with those identified in \citet{adami11} and \citet{willis13}.   At the significance and richness limit of SpARCs the overlap is primarily restricted to the most X-ray luminous systems at $z <$ 0.6, where the IR detection rate is low. Only one common cluster has a BCG detected at 24$\mu$m and  it is not unusual in its X-ray luminosity.   X-ray observations of the SpARCS sample  would provide a simple yet clear confirmation of this idea  -- the 20\% of the sample which is IR-luminous should be cool-core clusters when observed in the X-ray.

\section{Summary, Conclusions and Future Work}

 We investigate the infrared properties of a large sample (535) Brightest Cluster Galaxies in massive optically selected galaxy clusters over the redshift range $0.2 < z < 1.8$, using the SWIRE public 24$\mu$m imaging and catalogs.  We find several important results:
 
 \begin{itemize}
 \item{We detect 125/535, or 23\% of the BCGs at 24$\mu$m. The fraction of BCGs detected at 24$\mu$m above 100$\mu$Jy and above a fixed IR luminosity increases beyond redshift $z \sim$ 1.   Below $z < $ 1 no BCGs show L$_\mathrm{IR} > $ 10$^{12}$ L$_\odot$, whereas above $z > $ 1 this fraction rises to 20\%. }
 \item{An investigation of the  Spitzer-IRAC  IR colors  indicates that the bulk of the IR emission of  BCGs is not AGN dominated, but rather is due to dust enshrouded star formation.  Only 7/125 of the 24$\mu$m-detected BCG sample inhabit the AGN region  (defined by Donley et al.(2012)) of the IRAC color-color diagram.  }
 \item{Assuming no contribution from AGN to the infrared luminosity, we calculate the star-formation-rates  for the BCGs. These range from $\sim$1 M$_\odot$yr$^{-1}$ at low redshift to $\sim$ 1000M$_\odot$yr${-1}$ at high redshift.  At all redshifts, however, the specific SFRs of the BCGs are consistent with those of main-sequence star-forming galaxies of similar mass, and thus the bulk of the BCG population is not starbursting.  }
\item{We argue the the star formation episodes are short-lived and below $z \sim$ 1 do not contribute more than 10\% to the final total stellar mass of the BCGs.  In this case BCG growth is likely dominated by dry mergers. Above $z\sim$ 1 however, the star formation seen here may double the stellar mass of the BCGs and is thus an important, and perhaps dominate, process for stellar mass growth.}
\item{The physics driving the increase in star formation above  is as yet unconstrained, although the one object for which we have deep and high spatial resolution imaging (SpARCS1049+56) shows clear signs of wet merger activity.   }

\end{itemize}

\acknowledgments

Funding for PRIMUS is provided by NSF (AST-0607701, AST-0908246, AST-0908442, AST-0908354) and NASA (Spitzer-1356708, 08-ADP08-0019, NNX09AC95G). TMAW acknowledges the support of an NSERC Discovery Grant.
Funding for SDSS-III has been provided by the Alfred P. Sloan Foundation, the Participating Institutions, the National Science Foundation, and the U.S. Department of Energy Office of Science. The SDSS-III web site is http://www.sdss3.org/.
SDSS-III is managed by the Astrophysical Research Consortium for the Participating Institutions of the SDSS-III Collaboration including the University of Arizona, the Brazilian Participation Group, Brookhaven National Laboratory, Carnegie Mellon University, University of Florida, the French Participation Group, the German Participation Group, Harvard University, the Instituto de Astrofisica de Canarias, the Michigan State/Notre Dame/JINA Participation Group, Johns Hopkins University, Lawrence Berkeley National Laboratory, Max Planck Institute for Astrophysics, Max Planck Institute for Extraterrestrial Physics, New Mexico State University, New York University, Ohio State University, Pennsylvania State University, University of Portsmouth, Princeton University, the Spanish Participation Group, University of Tokyo, University of Utah, Vanderbilt University, University of Virginia, University of Washington, and Yale University. Financial support for this work was provided by NASA through programs GO-13306, GO-13677, GO-13747, GO-13845 \& GO-14327 from the Space Telescope Science Institute, which is operated by AURA, Inc., under NASA contract NAS 5-26555. 

%% in parentheses, after the keyword, but they will not be verified.

{\it Facilities:} \facility{Spitzer Space Telescope (MIPS; IRAC)},  \facility{Magellan (IMACS)}, \facility{CTIO}, \facility{CFHT}.

\clearpage

\LongTables
\begin{deluxetable*}{lrcrrrc}
\tablewidth{0pt}
\tablecaption{Infrared Detected BCGs \label{data_table}}
\tablehead{
\colhead{RA}             & \colhead{Dec } & \colhead{redshift\tablenotemark{a}} &
 \colhead{S$_{24{\mu}m}$  (mJy)} & \colhead{L$_\mathrm{IR}$ (10$^{11}$ L$_\odot$)\tablenotemark{b}} & \colhead{SFR (M$_\odot$yr$^{-1}$)} &   \colhead{Stellar Mass (10$^{11}$ M$_\odot$)}  }
%\colhead{} & \colhead{of BCGs\tablenotemark{a}} & \colhead{ of BCGs } & \colhead{spurious}  \\
%\colhead{ } & \colhead{in sample} & \colhead{with S$_{24{\mu}m}  > $ 100$\mu$Jy}     & \colhead{matches}  }

\startdata
%10:40:47.4  &  56:43:57 & {\bf 0.172} &   829.7$\pm$15.5 & 0.04$\pm$0.02 & 0.8$\pm$0.4  & XXX\\
%10:53:56.5  &  59:09:15 & {\bf 0.199} &   2556.6$\pm$17.7 & 0.13$\pm$0.06  &  2$\pm$1 & XXX\\
10:47:22.8  &  57:00:51 & 0.238 &    1500.4$\pm$15.5 & 0.71$\pm$0.36  &  12$\pm$6  & 0.9\\
10:33:34.0  &  58:14:35 & {\bf 0.430} &   17900.7$\pm$28.6 & 16.1$\pm$1.2  &  281$\pm$218  & 1.0\\
10:49:22.3  &  56:56:43 & {\bf 0.244} &   1429.9$\pm$16.7 & 1.9$\pm$1.3 &  33$\pm$24  & 2.4\\
10:48:50.3  &  56:09:08 &   {\bf 0.468} & 187.0$\pm$17.3  &  0.34$\pm$0.24 & 6$\pm$4 & 1.2\\
10:42:24.6  &  57:59:49 & 0.387 &    627.3$\pm$16.5 & 1.1$\pm$0.8  &  19$\pm$13  & 0.6\\
10:39:30.1  &  57:14:35 & 0.425 &   386.9$\pm$18.6 & 0.94$\pm$0.58 &  17$\pm$10  & 0.6\\
10:50:56.6  &  57:07:20 & 0.475 &    221.4$\pm$12.2 & 0.66$\pm$0.40 & 12$\pm$7  & 0.6\\
10:37:38.9  &  58:45:29 & 0.688 &    329.0$\pm$16.8 &  2.2$\pm$0.7 &  39$\pm$12  & 1.8\\
10:51:57.3  &  56:59:05 & 0.700 &    638.7$\pm$12.8 & 4.4$\pm$1.4 &  77$\pm$24  & 1.0\\
10:44:48.9  &  59:12:16 & 0.737 &    239.0$\pm$12.7 & 1.9$\pm$0.5  &  32$\pm$8  & 1.0\\
10:39:46.1  &  58:54:17 & 0.750 &    1380.2$\pm$17.1 & 11.0$\pm$0.4  &  192$\pm$72 & 2.8\\
10:42:28.6  &  57:51:24 & 0.778 &    464.0$\pm$17.6 & 4.1$\pm$1.1 &  71$\pm$19  & 1.0\\
10:37:25.8  &  59:02:12 & 0.813 &    248.5$\pm$13.5 & 2.3$\pm$0.6 &  39$\pm$10  & 1.4\\
10:42:36.0  &  58:18:16 & 0.825 &    163.3$\pm$14.8 & 1.5$\pm$0.4 &  26$\pm$7  & 1.0\\
10:40:05.0  &  59:16:26 & 0.825 &    266.7$\pm$16.9 & 2.5$\pm$0.7  & 43$\pm$12  & 1.4\\
10:35:36.5  &  57:50:35 & 0.875 &    175.4$\pm$15.8 & 1.7$\pm$0.6  &  29$\pm$10  & 1.1\\
10:58:24.5  &  57:27:26  & 0.888 &    234.1$\pm$15.5 & 2.3$\pm$0.8  &  41$\pm$14  & 1.3\\
10:46:43.4  &  58:21:20  & 0.900 &    411.5$\pm$13.9 & 4.5$\pm$1.5  &  78$\pm$26  & 2.4\\
10:48:17.1  &  55:52:18 & 0.900 &    785.9$\pm$16.5 & 9.1$\pm$3.5  & 159$\pm$62 & 1.0\\
10:44:05.0  &  58:59:04 & 0.925 &    514.7$\pm$11.9 & 6.2$\pm$2.2  &  109$\pm$39  & 1.4\\
10:51:55.3  &  57:01:09 & 0.937 &    269.5$\pm$11.2 & 3.2$\pm$1.2 &  56$\pm$20  & 0.6\\
10:55:56.3  &  57:10:37 & 0.950 &    560.9$\pm$17.5 & 7.4$\pm$2.9 &  130$\pm$50  & 0.8\\
10:34:05.4  &  58:11:31 & 0.963 &    523.4$\pm$18.1 & 7.2$\pm$2.9  &  126$\pm$51  & 1.9\\
%10:47:33.5  &  56:24:34 & 0.963 &    287.5$\pm$17.9 & 3.8$\pm$1.6  &  66$\pm$28  & XXX\\
10:52:22.7  &  56:55:44 & 0.963 &    525.8$\pm$11.6 & 7.3$\pm$2.8  &  127$\pm$49  & 1.1\\
%10:44:13.7  &   59:50:37 & 0.975 &    285.4$\pm$15.7 & 3.9$\pm$1.6  &  68$\pm$29  & XXX\\
10:46:25.0  &  57:04:29 & 1.000 &    189.6$\pm$15.2 & 2.7$\pm$1.3  &  47$\pm$22  & 1.7\\
%10:43:13.6  &  60:14:56 & 1.006 &    654.7$\pm$22.1 & 10.8$\pm$5.3  &  189$\pm$92  & XXX\\
10:47:05.7  &  58:05:42 & 1.013 &    2551.0$\pm$19.7 & 54.1$\pm$20.3  &  947$\pm$356   & 2.6\\
10:52:06.0 &  56:56:30 & 1.038 &    485.5$\pm$11.8 & 8.8$\pm$3.8  &  153$\pm$66  & 1.7\\
10:47:49.7  &  57:43:37 & {\bf 1.029} &  388.9$\pm$14.1 & 6.9$\pm$2.8  &  120$\pm$50  & 1.0\\
10:47:38.8  &  57:52:29 & 1.075 &    1281.6$\pm$18.0 & 32.5$\pm$12.2  &  570$\pm$213  & 2.0\\
10:59:55.7  &  57:45:56  & 1.075 &    431.0$\pm$17.9 & 8.9$\pm$3.9  &  155$\pm$69  & 1.2\\
10:52:30.0  &  58:19:43 & 1.125 &    357.4$\pm$15.5 & 8.6$\pm$3.7  &  150$\pm$65  & 1.6\\
10:46:26.4  &  57:32:30 & 1.125  &    926.6$\pm$16.0 & 27.0$\pm$10.1  & 472$\pm$177  & 1.5\\
10:59:14.9  &  57:32:40 &  1.200 &    226.4$\pm$14.4 & 6.6$\pm$3.1 & 116$\pm$54  & 1.4\\
10:52:08.4  &  57:02:38 & 1.250 &    377.2$\pm$11.1 & 15.6$\pm$0.7  &  272$\pm$123 & 2.2\\
%10:36:10.7  &  59:10:24 & 1.425 &    838.4$\pm$35.3 & 68.0$\pm$23.0  &  1188$\pm$425  & XXX\\
10:50:07.2  &  57:16:51 &  1.600 &    662.7$\pm$12.3 & 82.2$\pm$24.3  &  1438$\pm$426  & 3.6\\
10:44:59.5  &  57:52:07 & 1.650 &    281.1$\pm$17.0 & 34.2$\pm$9.0  &  599$\pm$152  & 5.3\\
10:49:22.6  &  56:40:33 & {\bf 1.70} &    606.1$\pm$18.0 & 104$\pm$29  &  1819$\pm$511  & 4.5\\ \hline
 02:19:41.8  & -04:00:33.3  & {\bf 0.141 }  & 358.1$\pm$18.4 & 0.12$\pm$0.06  &  2$\pm$1  & 0.5 \\
 02:25:24.9   &  -03:47:35.4      &0.250     & 263.8$\pm$16.3 & 0.18$\pm$0.09  &  3$\pm$2  & 0.3\\
       02:24:26.9  &    -05:36:32.2      &0.262  &       1814.7$\pm$17.8& 1.1$\pm$0.5  &  19$\pm$9  & 0.7\\
      02:24:33.1    &  -04:53:56.1      &0.337      &   262.34$\pm$19.7& 0.34$\pm$0.29 &  6$\pm$5 & 0.5\\
       02:19:10.5    &  -03:43:34.4     &     {\bf 0.763} &  2684.4$\pm$18.8& 18.8$\pm$7.6  &  329$\pm$132  & 1.6\\
       02:20:13.3    &  -06:00:54.9      &0.688      &    339.5$\pm$20.9& 2.3$\pm$0.8  &  40$\pm$13  & 1.1\\
       02:20:27.8    &  -05:47:24.9     & 0.762     &     236.0$\pm$16.4& 1.9$\pm$0.5 &  34$\pm$9  & 1.2\\
       02:22:13.0    &  -04:21:58.4    & 0.788     &    318.1$\pm$20.2& 2.8$\pm$0.8 &  48$\pm$13  & 0.9\\
       02:24:24.0     & -02:58:02.7      &0.788      &   216.3$\pm$19.6& 1.9$\pm$0.5  &  33$\pm$9  & 0.9\\
       02:19:29.0     & -04:07:00.2     & 0.813       &  1086.4$\pm$20.4& 10.3$\pm$3.8  &  180$\pm$66  & 1.6\\
       02:18:34.4     & -05:00:43.6    & {\bf     0.651 } &  368.8$\pm$19.0& 3.5$\pm$1.0  &  61$\pm$17  & 1.7\\
       02:15:43.9     & -04:24:53.5     & 0.850      &   694.6$\pm$18.2& 7.$\pm$2.0  &  122$\pm$37  & 2.7\\
       02:24:29.2     & -04:10:13.1     & 0.900      &   1082.9$\pm$19.8& 12.4$\pm$5.4  &  235$\pm$95  & 1.3\\
       02:25:06.9      &-04:47:18.4     & 0.925        &  276.0$\pm$19.4& 3.2$\pm$1.2  &  55$\pm$21  & 1.8\\
       02:02:08.5      &-03:41:26.8     & 0.925      &  345.9$\pm$19.3& 4.0$\pm$1.5  &  70$\pm$27  & 1.0\\
       02:14:38.2     & -03:37:38.3     & 0.931      &   2241.6$\pm$22.2& 35.1$\pm$12.8  &  614$\pm$224  & 2.7\\
       02:23:05.8      &-04:13:35.5     & {\bf    1.048 }  & 240.1$\pm$18.2& 3.1$\pm$1.3  &  53$\pm$23  & 0.9\\
       02:20:54.8     & -03:32:57.5     & 0.994       & 1137.5$\pm$21.7& 20.9$\pm$8.8  &  356$\pm$153  & 3.3\\
       02:22:36.5      &-03:50:30.3     & {\bf   0.822} &  1039.2$\pm$19.7& 19.3$\pm$8.4  &  338$\pm$146  & 2.0\\
       02:16:38.2      &-03:28:44.6     & 1.00        &  698.5$\pm$21.1& 11.3$\pm$5.5 &  198$\pm$95  & 2.0\\
       02:22:54.3      &-04:14:11.9      & 1.00     & 362.3$\pm$21.3& 5.5$\pm$2.4  &  96$\pm$42  & 1.0\\
%       02:22:17.9      &-05:55:53.7     & 1.05       & 268.1$\pm$17.5& 4.7$\pm$2.1  &  83$\pm$37  & XXX\\
       02:16:03.2      &-03:33:57.9    & 1.05    &     352.5$\pm$16.4& 6.5$\pm$2.7  &  113$\pm$47  & 1.8\\
       02:18:05.2     &-05:00:10.5     & {\bf   1.095 } &  247.1$\pm$15.9& 4.8$\pm$2.1  & 83$\pm$36  & 2.2\\
       02:16:52.3    & -03:37:56.7  & 1.350        &  508.8$\pm$20.0& 32.0$\pm$11.7  &  559$\pm$204  & 2.3\\
       02:26:15.6     & -04:56:28.0     & 1.375        &  182.9$\pm$19.7 & 10.3$\pm$4.9  &  180$\pm$87  & 2.3\\
       02:21:43.8     & -03:21:57.0  &   1.375        & 1023.9$\pm$21.0& 72.2$\pm$27.1  &  1263$\pm$474  & 3.8\\
       02:27:30.5     & -04:32:03.4   &   1.450         & 435.9$\pm$18.2& 35.8$\pm$11.1  &  628$\pm$195 & 1.8 \\ 
   02:18:45.2   & -05:42:56.8 & 1.45 & 313.3$\pm$19.6& 24.8$\pm$8.3 & 434$\pm$145 & 2.0 \\ \hline  
       16:05:44.6  &     54:57:16.2  & 0.188    &    378.6$\pm$16.8 & 0.01$\pm$0.06  &  2$\pm$1 & 0.7 \\
      16:06:27.9     & 54:56:14.7   &0.188       & 231.0$\pm$12.9& 0.08$\pm$0.04 &  1$\pm$0.7  & 0.3 \\
      16:13:32.9     & 56:17:47.5   &0.212      &  683.5$\pm$17.4& 0.27$\pm$0.13  &  5$\pm$2  & 0.8 \\
      16:18:38.3      &55:17:13.7   &0.250     &  107.5$\pm$13.2& 0.08$\pm$0.04  &  1$\pm$1  & 0.9 \\
      16:14:51.5     & 54: 02: 22.0  & 0.275     & 1042.6$\pm$16.7& 0.72$\pm$0.36  &  13$\pm$6  & 0.5 \\
      16:19:14.8      &55:13:40.8  & 0.313     &  224.6$\pm$16.4& 0.21$\pm$0.21  &  5$\pm$4  & 0.7 \\
      16:10:11.2     & 53:46:24.5  & 0.362     &  290.0$\pm$15.9& 0.44$\pm$0.37  &  8$\pm$7  & 0.9 \\
      16:13:17.5     & 56:01:19.3  & 0.387     & 2571.9$\pm$18.8& 3.9$\pm$2.8  &  68$\pm$49  & 1.3 \\
      16:00:37.2     & 55:27:26.9   &{\bf 0.465 }  &  163.5$\pm$16.2& 0.33$\pm$0.23  &  6$\pm$4  & 1.4 \\
  %    16:16:12.1    &  56:26:54.7   &0.575      &  242.1$\pm$17.4& 1.2$\pm$0.5  &  22$\pm$9  & XXX \\
      16:08:57.2     & 56:00:18.6   &0.688      & 524.1$\pm$17.0& 3.5$\pm$1.1  &  61$\pm$20  & 7.2 \\
      16:02:32.3     & 54:56:59.6   &0.700       & 150.0$\pm$14.5& 1.1$\pm$0.3  &  18$\pm$6  & 0.8 \\
      16:18:40.1    &  54:50:48.0   &0.700       & 759.9$\pm$15.1& 5.2$\pm$1.7  &  91$\pm$29  & 1.2 \\
      16:06:28.8     & 55:33:07.7   &0.725      & 165.3$\pm$16.1& 1.2$\pm$0.3  &  22$\pm$6  & 0.9 \\
      16:18:26.4    &  54:58:28.7   &0.737      &  278.2$\pm$13.0& 2.1$\pm$0.6  &  38$\pm$10  & 1.3 \\
      16:11:01.3     & 54:17:05.3   &0.750       & 695.7$\pm$15.7& 5.5$\pm$1.6  &  97$\pm$28  & 1.8 \\
      16:03:04.5     & 54:57:24.3   &0.775       & 607.1$\pm$15.9& 5.2$\pm$1.4  &  90$\pm$25 & 1.2 \\
      16:09:29.4     & 54:29:40.6   &0.788     &1764.0$\pm$16.7& 17.3$\pm$6.0 &  304$\pm$106  & 0.8 \\
      16:17:29.9    &  55:58:18.2   &0.788      & 207.1$\pm$16.2& 1.8$\pm$0.5  &  31$\pm$8  & 1.0 \\
      16:13:05.7     & 55:59:50.0  & 0.825       & 169.9$\pm$13.8& 1.5$\pm$0.4  & 27$\pm$7  & 1.4 \\
      16:07:16.8     & 55:32:59.8   &0.825     &  600.2$\pm$13.8& 5.7$\pm$1.6  &  100$\pm$28  & 1.1 \\
      16:08:25.6     & 54:45:08.9   &{\bf  0.906 }   & 610.6 $\pm$13.4& 5.8$\pm$1.6  &  102$\pm$29  & 0.8 \\
      16:19:19.7    &  55:36:45.7   &0.862      & 299.8$\pm$13.9& 2.9$\pm$0.9  &  51$\pm$15  & 0.6 \\
      16:02:50.7     & 54:54:52.2  & 0.862       & 337.9$\pm$14.7& 3.3$\pm$1.0  &  58$\pm$17  & 0.9 \\
      16:12:10.4     & 56:07:53.6   &0.888      & 483.1$\pm$13.9& 5.2$\pm$1.7 &  90$\pm$29  & 0.8 \\
      16:11:12.7     & 55:08:23.6   &{\bf 0.907 }  & 3065.4$\pm$19.9& 43.7$\pm$15.8  &  766$\pm$275  & 1.8 \\
      16:08:07.8      &54:19:43.1   &0.900       & 765.5$\pm$17.5& 8.8$\pm$3.3  &  154$\pm$57  & 1.4 \\
      16:18:18.0     & 54:35:38.9   &0.944      &  671.7$\pm$17.0& 8.8$\pm$3.7  &  155$\pm$66  & 1.8 \\
      16:22:42.1     & 54:50:55.0   &1.000       & 390.7$\pm$19.7& 6.0$\pm$2.6  &  105$\pm$45  & 1.1 \\
      16:02:43.5     & 54:49:09.7   &1.013        &194.3$\pm$17.4& 2.9$\pm$1.4  &  51$\pm$24  & 0.8\\
      16:13:01.7     & 54:46:10.0   &{\bf  1.092}    &1220.8$\pm$14.5& 24.2$\pm$9.5  &  423$\pm$167  & 1.8 \\
      16:08:27.1     & 54:36:47.3   &1.125        &345.3$\pm$14.0& 8.3$\pm$3.3  &  145$\pm$58  & 1.2 \\
      16:07:25.7     & 54:40:41.2   &1.125        &250.6$\pm$15.0& 5.8$\pm$2.5  &  101$\pm$43  & 0.7 \\
      16:11:15.8     & 54:15:11.4   &1.125       & 365.7$\pm$15.6& 8.8$\pm$3.9  &  153$\pm$69  & 1.3 \\
      16:05:55.2     & 55:31:47.2   &1.350       & 326.0$\pm$14.2& 19.5$\pm$8.0  &  342$\pm$138  & 2.6 \\
      16:06:52.7     & 55:39:36.6   &1.650       & 241.9$\pm$16.5& 28.3$\pm$7.3  &  496$\pm$128  & 4.6 \\ \hline
%16:32:33.1  &     41:32:17.3   &0.438       & 942.6$\pm$20.7 & 2.1$\pm$1.3 &    37$\pm$22 & XXX\\
      16:40:33.2 &     41:23:11.7   &0.662     & 220.6$\pm$16.2& 1.4$\pm$0.5  &  25$\pm$8  & 1.6 \\
      16:34:35.3   &   41:36:14.6   &{\bf 0.781}  &   584.5$\pm$15.1& 4.2$\pm$1.3  &  73$\pm$22  & 1.3 \\
      16:38:22.9     & 41:53:19.1   &0.862    &772.1$\pm$16.4& 8.0$\pm$2.5  &  141$\pm$44  & 1.3 \\
      16: 37:35.1     & 40:17:18.1  & 0.912      & 217.1$\pm$16.3& 2.3$\pm$0.9 &  40$\pm$16  & 2.9 \\
      16:41:52.4     & 41:34:01.5   &0.925        & 297.8$\pm$17.8& 3.4$\pm$1.3  &  60$\pm$23  & 0.7 \\
      16:37:49.5     & 42:04:55.3   &0.931      & 2767.4$\pm$18.5& 43.7$\pm$16.1  &  765$\pm$281  & 3.7 \\
      16:37:01.1     & 41:31:02.4   &0.944       & 747.8$\pm$17.1& 10.0$\pm$4.5  &  175$\pm$78  & 1.3 \\
      16:37:46.5     & 40:37:32.3   &0.944      & 507.1$\pm$17.4& 6.5$\pm$2.5  &  114$\pm$44  & 2.3 \\
      16:31:06.1     & 40:50:31.3   &0.956       & 340.8$\pm$17.8& 4.3$\pm$1.8  &  76$\pm$31  & 1.1 \\
      16:38:07.1     & 41:57:34.2   &1.075        &362.9$\pm$16.4& 7.3$\pm$3.0  &  128$\pm$53  & 0.8 \\
      16:40:22.3     & 42:03:14.4   &1.125        &295.8$\pm$13.3& 7.0$\pm$2.9 &  122$\pm$50  & 1.2 \\
      16:37:27.0     & 41:57:00.6   &1.125        &334.7$\pm$13.3& 8.0$\pm$3.2 &  139$\pm$57  & 1.3 \\
      16:41:40.3     & 41:13:18.4  & 1.125      & 283.4$\pm$15.0& 6.4$\pm$2.8  &  116$\pm$49  & 1.3 \\
      16:39:12.2     & 40:57:25.4  & 1.275    &    269.8$\pm$16.1& 10.9$\pm$5.7  &  190$\pm$100  & 1.7 \\
      16:37:59.2     & 42:03:56.5  & 1.575    &    175.0$\pm$12.9& 17.0$\pm$5.0  &  293$\pm$87  & 1.7 \\
      16:37:37.9     &41:17:28.6  &  1.775   &     339.4$\pm$17.9& 52.9$\pm$14.3  &  925$\pm$249  & 2.2 \\

%Lockman & 202 } &  43   & 2 \\
%ELAIS-N1   & 145  & 36 &  2 \\
%ELAIS-N2   & 64 & 17  & 1 \\
%\hline
%total & 535 & 125 & 7 \\

 \enddata
\tablenotetext{a}{Bold denotes a spectroscopic redshift, non-bold corresponds to the red sequence estimated redshift. } 
\tablenotetext{b}{LIRs estimated using the 24$\mu$m flux and \citet{chary01} methodology.}
%make this footnote extra long so that it extends over two lines.}
%% You can append references to a table using the \tablerefs command.
%\tablerefs{
%(1) Barbuy, Spite, \& Spite 1985; (2) Bond 1980; (3) Carbon et al. 1987;}
\end{deluxetable*}

\end{document}